\documentclass[11pt]{article}

\usepackage[margin=1in]{geometry}
\usepackage{amsmath,amssymb,bm}
\usepackage{graphicx}
\usepackage{booktabs}
\usepackage{enumitem}
\usepackage{comment}
\PassOptionsToPackage{bookmarks=false}{hyperref} 
\usepackage{hyperref}
\usepackage{authblk}  
\hypersetup{colorlinks=true,linkcolor=blue,citecolor=blue,urlcolor=blue}

\newcommand{\Sdata}{\mathcal{S}_{\mathrm{data}}}
\newcommand{\Rdata}{\mathcal{R}_{\mathrm{data}}}
\newcommand{\Rtrue}{\mathcal{R}_{\mathrm{true}}}
\newcommand{\Bdata}{\mathcal{B}_{\mathrm{data}}}
\newcommand{\Btrue}{\mathcal{B}_{\mathrm{true}}}
\newcommand{\Pdata}{\mathcal{P}_{\mathrm{data}}}
\newcommand{\Oop}{\mathcal{O}}

\begin{document}

\title{\vspace{-1.0em}%
Comment on “Exploring Data-Driven Corrections for $\phi$-Meson Global Spin Alignment Measurements” (arXiv:2508.18409)}

\author[1]{Jinhui Chen}
\author[2]{Diyu Shen}
\author[2]{Xu Sun}
\author[3]{Aihong Tang}
\author[1]{Baoshan Xi}

\affil[1]{Key Laboratory of Nuclear Physics and Ion-beam Application (MOE), Institute of Modern Physics, Fudan University, Shanghai 200433, China }
\affil[2]{Institute of Modern Physics, Lanzhou, 509 Nanchang Road, Lanzhou 730000, China}
\affil[3]{Brookhaven National Laboratory, Upton, New York 11973, USA}

\maketitle

\begin{abstract}

The method in arXiv:2508.18409 constructs a “data-driven correction” from combinatorial (pseudo-$\phi$) pairs and applies it to the signal. An explicit decomposition shows that the construction calibrates the \emph{background} response rather than the \emph{signal}: it is defined by the difference between an acceptance-free pseudo-$\phi$ surrogate and its data-level realization. Promoting a background-derived correction to a signal correction requires a \emph{strong physics proof} that signal and background share identical detector response at the pair level—including acceptance–anisotropy couplings and dependencies on parent kinematics—which the manuscript does not establish. Consequently, local numerical proximity in a restricted region of phase space is incidental rather than evidentiary; validation must rest on mechanism, not numerical coincidence. Moreover, the pseudo-$\phi$ background is non-unique: with infinitely many admissible constructions, any apparent agreement for a few cases would not be dispositive—no finite scan can substitute for a mechanism-level response equivalence. In the absence of such a demonstrated equivalence, the construction should be regarded as a background calibration rather than a signal correction.

\end{abstract}

\section{Executive summary}

The central claim in \cite{PaperUnderDiscussion} is that a data-driven procedure can correct detector effects for the $\phi$-meson spin-alignment observable (e.g., a $\cos^2\theta^\ast$--based proxy). However, by construction, the procedure is tuned on \emph{pseudo-$\phi$} (combinatorial) pairs and thus provides a principled correction for the \emph{background class it is trained on}. Applying it to the signal requires an additional, unproven identification of background and signal detector responses at the pair level. Without that identification, the procedure cannot be promoted to signal correction. Local numerical proximity in a restricted region of phase space has no standing as a general validation.

\section{Definitions and notation}
We consider a selected invariant-mass window in which the recorded (data-level) ensemble contains:
\begin{itemize}[leftmargin=1.2em]
  \item real $\phi$ decays (signal), denoted $\Rdata$;
  \item combinatorial $K^+K^-$ pairs falling in the $\phi$ window (pseudo-$\phi$ background), denoted $\Bdata$;
  \item an additional residual class of inclusive ``primary'' pairs (or other non-$\phi$ backgrounds), denoted $\Pdata$.
\end{itemize}
Let $y$, $k$, and $(1-y-k)$ be their corresponding \emph{data-level} fractions in the same selection. The mixture relation for the observable $\Oop[\cdot]$ (e.g., $\langle \cos^2\theta^\ast\rangle$ or a linear $\rho_{00}$ proxy\footnote{All statements below hold identically under the standard linear mapping between $\langle\cos^2\theta^\ast\rangle$ and $\rho_{00}$; because the map is linear, the algebra is unchanged.}) is
\begin{equation}
  \Oop[\Sdata] \;=\; y\,\Oop[\Rdata] \;+\; k\,\Oop[\Bdata] \;+\; (1-y-k)\,\Oop[\Pdata].
  \label{eq:mixture}
\end{equation}
We write $\Rtrue$ and $\Btrue$ for the corresponding \emph{truth-level} (acceptance-free) signal and pseudo-$\phi$ background ensembles. This relation fixes the notation and will be referenced for clarity below; it does not enter any derivations or fits.

\section{Algebraic decomposition of the correction construction}
We now define two internally labeled constructs that mirror the ingredients used by \cite{PaperUnderDiscussion}; these labels are introduced here for clarity.

\paragraph{Group 1 (background template; authors’ “Data Folding”, DF).}
A combinatorial template is built from rotated/mixed events to represent pseudo-$\phi$ background at the acceptance-free level (``truth'' surrogate) :
\begin{equation}
  \Oop[\text{Group 1}] \equiv \Oop[\Btrue].
  \label{eq:G1}
\end{equation}

\paragraph{Group 2 (inclusive selection and its mixed–weighted realization; authors’ “Data Scaling”, DS).}
The data-level selection within the $\phi$ window is the inclusive mixture
\begin{equation}
  \Oop[\text{Group 2}] \equiv \Oop[\Sdata]
  \;=\; y\,\Oop[\Rdata] + k\,\Oop[\Bdata] + (1-y-k)\,\Oop[\Pdata].
  \label{eq:G2}
\end{equation}
From this same selection, a \emph{mixed, weighted} realization is formed—authors’ “Data Scaling” (DS)—that, by design, isolates the \emph{data-level} pseudo-$\phi$ background:
\begin{equation}
  \Oop[\text{Group 2, mixed, weighted}] \;=\; \Oop[\Bdata].
  \label{eq:G2mw}
\end{equation}
Operationally, per-track weights in $(p_T,\eta,\phi-\psi_2)$ are constructed by comparing the decay-kaon and inclusive single-kaon distributions, and $\langle\cos^2\theta^\ast\rangle$ is then evaluated on rotated/mixed $K^+K^-$ pairs built from the weighted tracks; this enforces daughter-level kinematic matching and yields the right-hand side of Eq.~\eqref{eq:G2mw}.

\paragraph{The difference labeled as ``correction''.}
The inclusive selection (Group~2) is at the data level and thus carries the detector response; its mixed, weighted variant isolates the data-level pseudo-$\phi$ component, $\Oop[\Bdata]$. The background template (Group~1) is constructed to serve as a truth surrogate for pseudo-$\phi$, $\Oop[\Btrue]$. The intended logic is therefore to estimate the background detector response by the difference
\begin{equation}
  \Delta \;\equiv\; \Oop[\text{Group 1}] - \Oop[\text{Group 2, mixed, weighted}]
  \;=\; \Oop[\Btrue] - \Oop[\Bdata].
  \label{eq:DeltaDef}
\end{equation}
and then to apply to the signal, with the aim of removing detector effects. Note that, \emph{by construction}, $\Delta$ is \emph{purely} a background (pseudo-$\phi$) response difference.

\paragraph{Non-uniqueness of the pseudo-$\phi$ construction.}
The pseudo-$\phi$ background is not unique. Let $T$ denote any transformation that destroys true-parent
correlations while preserving the relevant single-kaon kinematics (e.g., event mixing; random rotations; two-dimension and three-dimension rotations; fixed
$\Delta\varphi$ rotations; inter-event shuffles ...). Denote by $\Bdata^{(T)}$ and $\Btrue^{(T)}$ the data-level and truth-surrogate
backgrounds produced by $T$. The constructed quantity is then
\[
\Delta^{(T)} \equiv \Oop\!\big[\Btrue^{(T)}\big] - \Oop\!\big[\Bdata^{(T)}\big].
\]
Because each transformation carries its own kinematic characteristics, $\Delta$ depends on the choice of $T$. Different, equally legitimate background realizations generally yield
different $\Delta^{(T)}$, underscoring that apparent agreement for a subset of choices cannot by itself establish applicability to the signal.

\paragraph{Necessary condition to apply a background-built correction to the signal.}
The quantity of interest is the truth-level signal observable $\Oop[\Rtrue]$. To justify using a background-built $\Delta$ as a \emph{signal} correction, one must assume an identity of detector responses between signal and pseudo-$\phi$ background:
\begin{equation}\label{eq:ResponseIdentity}
  \underbrace{\Oop[\Rdata]-\Oop[\Rtrue]}_{\text{signal response}}
  \;=\;
  \underbrace{\Oop[\Bdata]-\Oop[\Btrue]}_{\text{background response}} \, .
\end{equation}
In the absence of this identity, the procedure in Ref.~[1] is not a valid signal correction. Moreover, Eq.~\eqref{eq:ResponseIdentity} is not a generic truth: signal pairs arise from a common parent with two-body kinematics and characteristic pairwise correlations, whereas combinatorial pairs do not, and acceptance–anisotropy couplings act differently in these classes. To enforce an equal response, the authors restrict the pseudo-$\phi$ sample in invariant mass, implicitly hoping that at fixed $m_{\mathrm{inv}}$ the detector effects match those of real $\phi$. However, the detector response depends on the full parent four-momentum and on daughter/pair kinematics; matching only $m_{\mathrm{inv}}$ is far from sufficient to guarantee equality. Consequently, any numerical proximity between a background-built correction and the signal is at most incidental; apparent agreement without mechanism is coincidence, not confirmation.

\section{First closure test: acceptance stability under $|\eta|$ cuts}
Finite acceptance is a leading driver of detector distortion for decay-angle observables. Even at vanishing $v_2$, $|\eta|$ cuts and $p_T$-dependent efficiencies bias $\cos^2\theta^\ast$--type distributions; with anisotropy, acceptance--flow couplings further complicate the response. A credible closure must therefore show that the \emph{final corrected} observable is \emph{stable} under realistic acceptance changes.

\paragraph{Validation study.} 
Evaluate closure at $|\eta|<0.5$ and $|\eta|<1.0$ for both real-$\phi$ and pseudo-$\phi$, using rotated and mixed backgrounds. The final corrected values should be consistent within reasonable uncertainties across these acceptances. Note that this closure test is a self-consistency check and, by itself, does not establish the legitimacy of the correction. Even if this test passes, Eq.~(\ref{eq:ResponseIdentity}) still needs to be established; otherwise the construction remains a background calibration.

\section{``Background scanning'' cannot substitute for a signal correction}
The idea that scanning among backgrounds might ``cover'' the signal correction ``by chance'' is not sufficient to establish applicability to the signal. The pseudo-$\phi$ background is non-unique: there is an infinite family of admissible constructions (rotated, mixed, random or fixed-angle rotations, two-dimension and three-dimension rotations, shuffles etc.) that remove true-parent correlations while preserving single-particle kinematics. Each such background recipe carries its own pair-level kinematic imprint and acceptance couplings, so the resulting “correction” will, in general, differ across constructions.
 Consequently, apparent agreement for a few choices cannot be dispositive; a finite scan cannot substitute for a mechanism-level equivalence of detector responses.

\begin{itemize}[leftmargin=1.2em]
  \item \textbf{Signals are not backgrounds.} Signal pairs inherit two-body kinematics from a common parent; combinatorial pairs do not. 
  \item \textbf{No generic phase-space correlation.} There is no principle tying signal-specific acceptance biases to those of combinatorial backgrounds across the full phase space.
  \item \textbf{Spectral analogy.} In the spectra analysis for resonances, 
  one would not take an efficiency correction derived purposely from \emph{combinatorial background} and apply it to \emph{signal}. The same logic applies here.
\end{itemize}

\section{Conclusions}

We have examined the “data-driven correction” in Ref.~\cite{PaperUnderDiscussion}; a straightforward decomposition of the construction from pseudo-$\phi$ pairs and the inclusive selection shows that the operation calibrates the background response via the difference between an acceptance-free pseudo-$\phi$ surrogate and its data-level realization. Elevating this construction to a signal correction would require the response-equivalence identity of Eq.~(\ref{eq:ResponseIdentity}), which is neither generic nor established. Because pseudo-$\phi$ admits infinitely many admissible realizations, scanning over backgrounds cannot establish applicability, and local numerical proximity in restricted regions of phase space is incidental rather than evidentiary. In the absence of a mechanism-based justification—independent of implementation details—the method of Ref.~\cite{PaperUnderDiscussion} is best regarded as a background calibration rather than a signal correction. Ultimately, validation rests on mechanism, not on numerical proximity.

\end{document}